\documentstyle[12pt,epsfig]{article}
\normalbaselines
\catcode `\@=11
\@addtoreset{equation}{section}

\def\section{\@startsection {section}{1}{\z@}{-3.5ex plus -1ex minus
     -.2ex}{2.3ex plus .2ex}{\normalsize\bf}}
\def\subsection{\@startsection{subsection}{2}{\z@}{-3.25ex plus -1ex minus
 -.2ex}{1.5ex plus .2ex}{\normalsize\bf}}

\def\thebibliography#1{\section*{References\markboth
  {REFERENCES}{REFERENCES}}\list
  {[\arabic{enumi}]}{\settowidth\labelwidth{[#1]}\leftmargin\labelwidth
  \advance\leftmargin\labelsep
  \usecounter{enumi}}
  \def\newblock{\hskip .11em plus .33em minus -.07em}
  \sloppy
  \sfcode`\.=1000\relax}
 
\catcode `\@=12
\begin{document}
\vspace*{2.5cm}
\noindent
{ \bf BOGOMOL'NYI SOLITONS AND 
HERMITIAN SYMMETRIC SPACES }\vspace{1.3cm}\\
\noindent
\hspace*{1in}
\begin{minipage}{13cm}
Phillial Oh
\vspace{0.3cm}\\
Department of Physics, Sung Kyun Kwan  University\\
Suwon 440-746, Republic of Korea \\
\makebox[3mm]{ } and \\
Center for Theoretical Physics, M.I.T.\\
Cambridge, MA 02139-4307. U.S.A. \\
E-mail: ploh@ctpa03.mit.edu
\end{minipage}
\vspace*{0.5cm}
\begin{abstract}
\noindent
We apply the coadjoint orbit method to construct relativistic
nonlinear sigma models (NLSM) on the target space of
coadjoint orbits coupled with the Chern-Simons (CS) gauge field  and 
study self-dual solitons. When the target space is given by 
Hermitian symmetric space (HSS), 
we find that the system admits self-dual solitons whose energy is 
Bogomol'nyi bounded from below by a topological charge. 
The Bogomol'nyi potential 
on the Hermitian symmetric space is obtained in the case when the 
maximal torus subgroup is gauged, and the self-dual equation in the
$CP(N-1)$ case is explored. We also discuss the self-dual solitons
in the non-compact $SU(1,1)$ case and present a detailed analysis 
for the rotationally symmetric solutions.
\end{abstract}

\section{\hspace{-4mm}.\hspace{2mm} INTRODUCTION}

Recently, a coadjoint  orbit method to formulate  the
nonlinear sigma model  defined on the target space 
of homogeneous space $G/H$ was proposed \cite{oh1}.
It was first applied to a non-relativistic
spin system whose Poisson bracket 
between the dynamical variables
defined on the coadjoint orbit   
satisfies the classical ${\cal G}$ algebra.   
The Euler-Lagrange equation of motion 
yields the generalized continuous Heisenberg ferromagnet 
\cite{fad,lak}. 
When the target space of coadjoint orbit is given by HSS 
 which is a symmetric space equipped with  
complex structure \cite{ford}, 
the generalized ferromagnet system becomes   
completely integrable in 1+1 dimension \cite{oh1}.
Later, this method was exploited to produce a class of
integrable extension of relativistic NLSM
in 1+1 dimension \cite{ohh}. It was also discovered that
incorporation the CS gauge field in 2+1 dimension 
on the same target space 
produces  a class of self-dual field theories which  
admit Bogomol'nyi self-dual equations 
saturating the energy functional 
\cite{oh2}.  A detailed numerical investigation
in the compact $SU(2)$ case 
\cite{oh3} showed a rich structure of
self-dual solitons in the system.

In this paper, we apply the coadjoint orbit method to
construct relativistic NLSM on the target space of
coadjoint orbits coupled with the CS gauge field and 
study self-dual solitons. When the target space is HSS,
the Hamiltonian is bounded from below by 
a topological charge, and the resulting self-dual CS solitons  
satisfy a vortex-type equation, thus producing a class of
new self-dual theories on HSS. This construction provides a
unified framework for treating the previous  gauged $O(3)$ model on 
$S^2$ and $CP(N-1)$ models \cite{nard}  which are well known examples
of the coadjoint orbit $G/H$ with $S^2=SO(3)/SO(2)\approx SU(2)/U(1)$
and $CP(N-1)=SU(N)/SU(N-1)\times U(1)$. 
We also study the self-dual solitons in the non-compact HSS 
with $SU(1,1)$ group in which the target space is given by the 
upper sheeted hyperboloid and find various topological
and nontopological solitons.

We first give a brief summary of NLSM on the
target space  of coadjoint orbit for completeness.
Consider a group $G$, Lie algebra ${\cal G}$
and its dual ${\cal G}^*: X \in {\cal G};~~ u \in 
{\cal G}^*$. The adjoint action of $G$ on the Lie 
algebra is defined by
\begin{equation}
\mbox{Ad}(g)X=gXg^{-1},~~g\in G.
\end{equation}
Denoting inner product between ${\cal G}$
and ${\cal G}^*$ by $<u,X>$,  the coadjoint action of the group 
on ${\cal G}^*$ is defined in such a way to make the 
inner product invariant:
\begin{equation}
<\mbox{Ad}^*(g)u,X>=<u,\mbox{Ad}(g^{-1})X>.
\end{equation}
The coadjoint orbit is given by the orbit of coadjoint action 
of the group $G$: Fix a point $u\in {\cal G}^*$, then the orbit is 
generated by 
\begin{equation}
{\cal O}_u=\{x\vert x=\mbox{Ad}^*(g)u, g\in G\}.
\end{equation}
It can be shown that ${\cal O}_u\approx G/H$, where $H$ 
is the stabilizer of the point $u$.
  
Let us assume that the inner product is given by the trace:
$<u,X>=\mbox{Tr}(Xu)$. Then, ${\cal G}$
and ${\cal G}^*$ are isomorphic and the coadjoint
orbit can be parameterized by
\begin{equation}
Q=gKg^{-1}=Q^At^B\eta_{AB};~t^A, K\in{\cal G}~ 
(A=1,\cdots, \mbox{dim}~{\cal G}),
\label{definition}
\end{equation}
where $\eta_{AB}$ is the $G$-invariant metric given by
$\mbox{Tr}(t^At^B)=-\frac{1}{2}\eta^{AB}$ with $t^A$'s
being the generator of ${\cal G}$.  
The action for the NLSM on the target space of 
coadjoint orbit can be constructed as
\begin{equation}
S(g)=\epsilon\mbox{Tr}\int d^3x \partial_\mu Q\partial^\mu Q.
\label{intact}
\end{equation}
$\epsilon=+1$ for the compact case $-1$ for the non-compact case
\cite{note}.
Let us first choose the element $K$
to be the central element of the Cartan subalgebra of ${\cal G}$
whose centralizer in ${\cal G}$ is $H$. 
Then, for the HSS, we have
$J = \mbox{Ad}(K)$ acting on the coset 
is a linear map satisfying the complex
structure condition $J^{2} = -1$, which gives the useful identity
\cite{oh1}:
\begin{equation}
[Q,[Q,\partial_\mu Q]]=-\partial_\mu Q.
\label{use}
\end{equation}
This paper is organized as follows: In Section 2, starting
from a CS gauged action of (\ref{intact})  
on arbitrary HSS, we derive self-dual equations and Bogomol'nyi
potential. We give  explicit expressions in $CP(N-1)$ case.
In Section 3, we  deal with  non-compact minimal $SU(1,1)$ model
and discuss rotationally symmetric solutions in detail.
In Section 4, we give the conclusion.
 
\section{\hspace{-4mm}.\hspace{2mm} COMPACT MODEL}

Let us consider the following
CS gauged action of (\ref{intact}):
\begin{equation}
S_G= \int  d^3x\left[-\epsilon\frac{1}{2}\left(
D_\mu Q^AD^{\mu}Q^B\eta_{AB}\right) 
-W_G(Q^A)-\kappa\epsilon^{\mu\nu\rho}\mbox{Tr}
\left(\partial_\mu A_\nu A_\rho+\frac{2}{3}A_\mu A_\nu A_\rho
\right)\right],
\label{action}
\end{equation}
where the covariant derivative is defined by 
\begin{equation}
D_\mu Q=\partial_\mu Q+[A_\mu,Q],~~~A_\mu=A_\mu^A t^B\eta_{AB}.
\end{equation}
We assume that the potential is given by
\begin{equation}
W_G(Q^A)=\frac{1}{2}I^{AB}Q^AQ^B,
\end{equation}
where $I^{AB}$ is the symmetric tensor and its content will be
determined by the self-duality condition.
The equations of motion are given by
\begin{eqnarray}
D_\mu [Q,~D^\mu Q]+[\bar Q,Q]&=&0, ~~(\bar Q=I^{AB}
Q^At^B).\nonumber\\
\frac{\kappa}{2}\epsilon^{\mu\nu\rho}F_{\nu\rho}&=&
[Q,D^\mu Q].
\label{csqq}
\end{eqnarray}

We first treat the compact case with $\eta_{AB}=\delta_{AB}$. 
To study self-dual solitons, we bring the energy functional 
into Bogomol'nyi expression:
\begin{eqnarray}
E_G&=&
\int d^{2}x \left[ \frac{1}{2}\left((D_0Q^A)^2
+(D_{i}Q^A)^2\right)+W(Q^A) \right] 
\nonumber\\
&=&\int d^{2}x \left[\frac{1}{2}(D_0Q^A)^2+
\frac{1}{4}( D_{i}Q^A\pm\epsilon_{ij}
[Q,  D_{j}Q ]^A)^2\right.\label{bogomol}\\ 
& &\left. +W(Q^A)\pm \frac{1}{2}\epsilon_{ij}F_{ij}^AQ^A\right] 
\pm 4\pi T_G ,\nonumber
\end{eqnarray}
where the topological charge $T_G$ is given by
\begin{equation}
T_G =\frac{1}{8\pi}
\int d^2x [\epsilon_{ij}Q^A 
[\partial_iQ,\partial_jQ]^A-2\epsilon_{ij}\partial_i (Q^AA_j^A)].
\end{equation}
In deriving (\ref{bogomol}), we used the 
gauged version of (\ref{use}) where $\partial_\mu$ is replaced
by the covariant derivative $D_\mu$ \cite{oh2}.
Thus, the Hamiltonian is  bounded from below by the topological 
charge $T_G$ when the potential $W_G$ is chosen such that
\begin{equation}
W_G \pm \frac{1}{2}\epsilon_{ij}F_{ij}^AQ^A = 0 .
\end{equation}
Here, $F_{ij}^A$ is determined in terms of $Q^A$ by the
Gauss's law which is the time-component of (\ref{csqq}).

The minimum energy arises when the self-duality
equation is satisfied:
\begin{equation}
D_iQ=\mp \epsilon_{ij}[Q, D_jQ].
\label{selfdual}
\end{equation}
Consistency with the static equations of motion
(\ref{csqq}) forces
\begin{equation}
F_{ij}=0,~~~ A_0=\pm\frac{1}{\kappa}Q,
\end{equation}
which in turn puts the potential $W_G=0$ and $I^{AB}=0$.
Note that the gauge field can be chosen as a pure gauge
in this case and the contents of the
Bogomol'nyi solitons are precisely the two dimensional
instantons which were completely classified on each HSS 
\cite{pere}.

More interesting cases in which the system offers 
other solitons  arise when we gauge the subgroup $H$. 
We consider gauging the
maximal torus subgroup of $G$:
\begin{equation}
S_H=\int d^{3}x
\left[-\frac{1}{2}\left(D_\mu Q^AD^\mu Q^A\right)-
W_H(Q^A)+\frac{\kappa}{2}\epsilon^{\mu\nu\rho}
\partial_\mu A^a_\nu A^a_\rho \right].
\label{subaction}
\end{equation}
Here, the index 
$a=1,\cdots, \mbox{rank}~ G$
denotes the maximal Abelian subgroup.
Again, the content of the potential $W_H$
will be determined from the self-duality condition.

Using the Gauss's  law given by
\begin{equation}
\frac{\kappa}{2}\epsilon_{ij}F^a_{ij} =-[Q, D_0Q]^a,
\label{aagauss}
\end{equation}
we find that the energy functional satisfies
\begin{eqnarray}
E_H&=&
\frac{1}{2}\int d^{2}x\left[\left(D_0Q^A\pm \frac{1}{\kappa}
[Q,Q_H]^A\right)^2
+\frac{1}{2}(D_{i}Q^A\pm \epsilon_{ij}[Q,D_{j}Q]^A)^{2}\right]
\pm 4\pi T_H,\nonumber\\
T_H&=&\frac{1}{8\pi}\int d^2x[\epsilon_{ij}
Q^A[\partial_iQ, \partial_jQ]^A
-2\epsilon_{ij}\partial_i(Q^a_H A^a_j)],
\end{eqnarray}
when the Bogomol'nyi potential $W_H$ is chosen as
\begin{equation}
W_H=\frac{1}{2\kappa^2}([Q_H, Q]^A)^2
,~~ Q_H\equiv Q^a_Ht^a=(Q^a-V^a)t^a.
\label{ppte}
\end{equation}
Note that $V^a$'s are free parameters associated with the 
vacuum symmetry breaking \cite{kimm}. When the self-duality 
equations
\begin{equation}
D_iQ^A=\mp \epsilon_{ij}[Q, D_jQ]^A,~~~
D_0Q^A=\mp \frac{1}{\kappa} [Q,Q_H]^A,
\label{last}
\end{equation}
are satisfied, we see that the energy is 
saturated by the topological charge:
\begin{equation}
E_H = 4\pi \vert T_H\vert.
\end{equation}
The first order equation (\ref{last}) in the static case fixes
$A_0^a$ to be
\begin{equation}
A_0^a=\pm\frac{1}{\kappa}Q^a_H,
\end{equation}
which automatically solves the
Euler-Lagrange equations of motion of the action 
(\ref{subaction}) with
the potential given by (\ref{ppte}).

Let us examine (\ref{aagauss}) and (\ref{last})
 more closely in $CP(N-1)$ case.
We use the expression of $Q$ \cite{oh1}, 
\begin{equation}
Q=i\Psi\Psi^\dagger-i\frac{I}{N}
\label{spinfunction}
\end{equation}
where the column vector $\Psi$ can be expressed by the Fubini-Study 
coordinate $\psi_a$'s  ~($a=1,2,\cdots, N-1$):
\begin{equation}
\Psi=\frac{1}{\sqrt{1+\vert\psi\vert^2}}
\left(\begin{array}{c}
1\\ \psi_1\\ \vdots\\ \psi_{N-1}\end{array}
\right),
\label{slp}
\end{equation}
with $\vert\psi\vert^2=\vert\psi_1\vert^2+\cdots+
\vert\psi_{N-1}\vert^2.$
Using the complex notation;
 $z = x+iy, \bar z = x-iy$,
$A_z  = \frac{1}{2}(A_1 - iA_2), A_{\bar z} =
\frac{1}{2} (A_1 + iA_2)$,  and
$D_z = \frac{1}{2}(D_1-iD_2),
D_{\bar z} = \frac{1}{2}(D_1+iD_2)$, we obtain
an alternative expression of the self-duality equation, 
\begin{equation}
D_zQ=\mp i[Q, D_zQ].
\label{selfdualeq}
\end{equation}
With the above parameterization of $Q$, the self-duality 
equation (\ref{selfdualeq}) for the plus sign  
becomes a set of $N-1$ equations \cite{oh2},
\begin{equation}
D_-^{a} \equiv \partial_z+
\frac{i}{2}(A_z^1+\frac{1}{\sqrt{3}}A_z^2
+\cdots +\sqrt{\frac{2}{a(a-1)}}A^{a-1}_z+
\sqrt{\frac{2(a+1)}{a}}A^a_z),
\end{equation}
\begin{equation}
D_-^a\bar\psi_a=0.
\label{equa1}
\end{equation}
Similarly, for the minus sign, we have
\begin{equation}
D_+^a \equiv \partial_z-\frac{i}{2}(A_z^1+
\frac{1}{\sqrt{3}}A_z^2
+\cdots +\sqrt{\frac{2}{a(a-1)}}A^{a-1}_z+
\sqrt{\frac{2(a+1)}{a}}A^a_z),
\end{equation}
\begin{equation}
D_+^a\psi_a=0.
\label{equa2}
\end{equation}

We concentrate on the plus sign from here on.
With $\bar\psi_a=w_a\exp(i\phi_a)$,
we find that (\ref{aagauss}), (\ref{last}) 
and (\ref{equa1})
produce the following  new vortex-type equation: 
\begin{equation}
\nabla^2\log w_a+
\epsilon_{ij}\partial_i\partial_j\phi_a=
\sum_{i=1}^{a-1}\sqrt{\frac{1}{2i(i+1)}}
\Gamma^i+\sqrt{\frac{a+1}{2a}}\Gamma^a,
\label{oheq}
\end{equation}
where $\Gamma^a$ is given by
\begin{equation}
\Gamma^a=\frac{V^b-Q^b}{\kappa^2}
\left[\frac{2}{N}\delta^{ab}+d^{abc} Q^c-Q^a Q^b\right].
\end{equation}
We used the normalization: $\{\lambda^A, \lambda^B\}
=(4/N)\delta^{AB}I+2d^{ABC}\lambda^C.$ Also the Bogomol'nyi
potential (\ref{ppte}) can be expressed by
\begin{equation}
W_H=\frac{1}{2\kappa^2}(V^a-Q^a)(V^b-Q^b)
\left[\frac{2}{N}\delta^{ab}+d^{abc} Q^c-Q^a Q^b\right].
\end{equation}
Let us give an example in  the case of $CP(1)$. With 
$w_1=w, \phi_1=\phi, V^1=V, $ 
the above potential becomes 
\begin{equation}
W_H=\frac{1}{2\kappa^2}(V-Q^3)^2(1-(Q^3)^2),
\label{su2po}
\end{equation}
which is exactly the same as the potential in the $O(3)$ model
\cite{kimm}. Next, we  find that  (\ref{oheq})  becomes
\begin{equation}
\nabla^2\log w+
\epsilon_{ij}\partial_i\partial_j\phi=
\frac{1}{\kappa^2}\left[V-\frac{1-w^2}{1+w^2}\right]
\left[1-\left(\frac{1-w^2}{1+w^2}\right)^2\right]
\end{equation} 
A detailed numerical study of the above equation showed
that the equation has various kinds of rotationally symmetric
solitons solutions connected with
symmetric and broken phases, and
they are anyons carrying fractional angular momentum  \cite{kimm}. 
Similar results are expected in the more complicated
higher $CP(N)$ case, but a detailed study will be addressed elsewhere.

\section{\hspace{-4mm}.\hspace{2mm} NON-COMPACT $SU(1,1)$ SOLITON} 

In this section, we consider a non-compact HSS with
$\epsilon=-1$. 
We restrict to the $SU(1,1)$ group with $\eta_{AB}=(-,-,+)$.
The target space is given by the two-sheeted hyperboloid
$H=SU(1,1)/U(1)$.
Using the expression for the group element $g$ of (\ref{definition})
given by
\begin{equation}
g=\frac{1}{\sqrt{1-\vert\psi\vert^2}}
\left(\begin{array}{cc}
1 &\psi^*\\
\psi &1\end{array}\right)
\end{equation}
which satisfies $gMg^\dagger=M$ with $M=\mbox{diag}(1,-1)$, we have
(with $K=i \sigma^3/2$)
\begin{equation}
Q=\frac{i}{2(1-\vert\psi\vert^2)}
\left(\begin{array}{cc}
1 +\vert\psi\vert^2 &-2\psi^*\\
2\psi &-(1+\vert\psi\vert^2)
\end{array}\right). 
\end{equation}
We restrict to $\vert\psi\vert<1$, which corresponds to the
upper sheet of ${\cal M}=SU(1,1)/U(1)$. 
A couple of remarks at this
point concerning the ungauged case are in order. First, some 
soliton solutions associated with non-compact NLSM
were discussed in connection with the 
Ernst equation \cite{take}, which are not self-dual.
Secondly, using the above expression, one can check that there 
actually exist self-dual soliton solutions
which are analytic or anti-analytic as in the
compact case \cite{bela}, but the energy
and topological charge diverge at the boundary
$\vert\psi\vert=1$. Coupling with CS gauge field greatly 
improves the situation, because the gauge field effectively provides
a potential barrier to the boundary (see (\ref{effect}))
and prevents the system from diverging.

Again, with the parameterization $\bar\psi=w\exp(i\phi)$, we find 
the Bogomol'nyi potential (\ref{ppte}) and the self-dual
equation (\ref{oheq}) are 
produced as follows:
\begin{eqnarray}
W_H&=&
\frac{1}{2\kappa^2}\left[V-\left(\frac{1+w^2}{1-w^2}\right)\right]^2
\left[\left(\frac{1+w^2}{1-w^2}\right)^2-1\right]
\label{boghrt}\\
\nabla^2\log w&+&
\epsilon_{ij}\partial_i\partial_j\phi=
\frac{1}{\kappa^2}\left[V-\frac{1+w^2}{1-w^2}\right]
\left[\left(\frac{1+w^2}{1-w^2}\right)^2-1\right].
\label{eqmotion}
\end{eqnarray} 
Let us look for the rotationally symmetric solutions with  
the ansatz in the cylindrical coordinate  $(r,\theta)$ given  by 
\begin{eqnarray}\label{ansat}
w=\tanh \frac{f(r)}{2},\;\phi=n\theta, \;
A_i=\frac{\epsilon_{ji}x_j}{r^2}a(r).
\end{eqnarray}
Then, the Gauss's law and self-dual equation become
($^\prime=d/dr$)
\begin{eqnarray}
a^\prime(r)&=& (r/k^2) (-V + \cosh f(r))(1-\cosh^2 f(r)),
\nonumber\\
 r f^\prime(r)&=& (a(r) -n)\sinh f(r).
 \label{osci}
\end{eqnarray}
Now, the combined equation of motion in (\ref{eqmotion})
becomes an analogue of the one dimensional Newton's equation
for $r>0$, if we regard $r$ as ``time''
and $u(r)\equiv \log\tanh \frac{f(r)}{2}$ 
as the ``position" of the hypothetical particle 
with unit mass under a time-dependent friction,
$(1/r)u^\prime$, and an effective potential $V_{eff}$:
\begin{equation}
V_{eff}(u)=\frac{1}{2\kappa^2}\coth^2 u+\frac{V}{\kappa^2}\coth u.
\label{effect}
\end{equation}
The exerting force also includes an impact term
at $r=0$ due to $\epsilon^{ij}\partial_{i}\partial_{j}\phi=
\frac{n}{r}\delta(r)$ in (\ref{eqmotion}).

The inspection of the effective potential
suggests that  solitons are basically of two types;
the non-topological vortices with $n\neq 0$ (negative integer)
and the non-topological
solitons with $n=0$. In the former case, the ``particle"  
starts from $u = -\infty$, reaches a 
turning point where it stops, changes the direction,
and finally rolls down to $u = -\infty$.
In the latter, the ``particle" starting at some finite position,
either rolls down to $u = -\infty$ directly, or moves to a
turning point, changes the direction,
and rolls down to $u = -\infty$.
Let us look at the solutions more closely.
Near $r=0$, the condition for $A_i$ to be non-singular
forces $a(0)=0$. First, when $n\neq 0 $, we must have
$f(0)=0$. When $n= 0$, $\alpha\equiv f(0)$ can be arbitrary. 
The behavior of the solution near $r=\infty$ can be
also read off from the conditions $f^\prime(\infty)
=a^\prime(\infty)=0$; $\beta\equiv f(\infty)=0$ for arbitrary $\gamma
\equiv a(\infty)$ and $V$. Putting $f(r)=f_\infty r^l,
~ a(r)=\gamma+a_\infty r^s ~(l,s<0)$ near $r=\infty$, we find  
$l=\gamma-n, s=2\gamma+2-2n$ for $V\neq 1$. Since $l,s<0$, we have 
consistency condition $\gamma<n-1$.
When  $V= 1$, we have $l=\gamma-n, s=4\gamma+2-4n$ and
$\gamma<n-\frac{1}{2}$.
When  $\beta=\cosh^{-1}V$ for $V> 1$, $\gamma$ must be
equal to $n~(\neq 0)$. This solution which will show oscillatory
behavior before
it comes to rest does not exit.
Near $r=\infty$, we assume an exponential approach
$f(r)=\beta +f_\infty r^le^{-ar},~ a(r)=\gamma
+a_\infty r^se^{-br} ~(a,b>0)$. 
Then substitution leads to a contradictory output, $l=s$ and $l=s+1$. 
Power law approach with $a,b=0$ and $l,s<0$ also 
leads to a contradiction. In view of the Bogomol'nyi
potential (\ref{boghrt}), this excludes any solitons
in the broken vacuum with $V=\cosh f(\infty)$, and
all the solitons are in the symmetric phases.

Let us focus on the vicinity of $r=0$.
(a) $V\leq 1$, in which the effective
potential (\ref{effect}) is a monotonically decreasing function.
(a-i)  $n\neq 0$;
Trying power solutions  of the form
$f(r)=f_0r^p,~ a(r)=a_0r^q ~(p,q>0)$, we find
$p=-n,~q=2-2n$ for $V< 1$. Hence $n$ must be a negative integer.
When  $V= 1$, we have $p=-n,~q=2-4n$.
(a-ii) $n= 0, \alpha\neq 0$;
Let us try $f(r)=\alpha +f_0r^p,~ a(r)=a_0r^q ~(p,q>0)$.
We find $p=q=2$. We note that both $a_0$ and $f_0$ turns out to be
negative, so that the solution rolls down to $r=\infty$.
Climbing up at first and then rolling down the hill
solution does not exist.
When (b) $V> 1$, the effective potential (\ref{effect})
develops a pool with a local minimum at $f_m=\cosh^{-1}V$. 
(b-i) $n\neq 0$; 
The behavior is similar to  (a-i)
 except the fact  $a(r)$ passes the minimum at $r_m$
twice  in the process of  climbing up,
passing the turning point, and rolling down the hill to
its original position.
(b-ii)  $n= 0, \alpha\neq 0$; There are two cases. When
$\alpha <\cosh^{-1}V$, the solution, if exists, will behave
similarly with (b-i) except that it starts at some
finite point $\alpha$. However, it cannot
exist for the following reason; The initial ``velocity" of the
particle is given by $u^\prime(0)\propto f^\prime(0)=0~(
f^\prime(r)\propto r$ from (a-ii)). Hence the particle does not carry
enough kinetic energy to return to its starting
point in this dissipative system with conservative potential.
Note that when $n\neq 0$, even though the initial velocity is 
in general equal to $0$ except $n=-1$, the solutions are possible
because of the impact term at $r=0$. 
In the opposite case $\alpha
>\cosh^{-1}V$, it is
similar to (a-ii) and only rolling down the hill is permitted. 
A detailed numerical study given in Figs. 1 and 2 indeed confirms the 
existence of these solitons.

Note that there does not exist any topological
lump solutions, because $\pi_2({\cal M})=0$.
And the topological vortices does not exist, because
there is no bump in the effective potential where the particle 
can stop at the top.   
In the solutions, the magnetic
flux is given by $\Phi=2\pi \gamma$, and the energy is saturated
by the topological charge; $E=4\pi\vert T\vert=
2\pi\vert \gamma(1-V)\vert$. 
The system also carry non-vanishing angular momentum.
Let us define
\begin{equation}
J=\int d^2x \epsilon_{ij}x_iD_0Q^AD_jQ^B\eta_{AB}.
\end{equation}
A simple calculation using the Gauss's Law (\ref{aagauss})
(with plus sign in the right hand side due to
the $\epsilon$ factor), and self-dual equations
(\ref{last}) and (\ref{osci}), we find $J=\pi\kappa((\gamma-n)^2
-n^2)$. Thus the solitons in general carry
 a fractional angular momentum, representing anyons.  For the
 non-topological solitons, 
it is simply  $J=\pi\kappa\gamma^2.$

\section{\hspace{-4mm}.\hspace{2mm} CONCLUSION } 

We showed that the coadjoint orbit approach
for the relativistic NLSM coupled with
CS gauge field leads to a class of new self-dual field theories on the
target space of HSS  which  contain  the previous $O(3)$ and
$CP(N-1)$ models, and a new non-compact $SU(1,1)$
model. We also found an explicit expression of
the Bogomol'nyi potential when
the maximal torus subgroup is gauged, and showed that the
non-compact NLSM  admits self-dual soliton solutions which
are saturated by the Bogomol'nyi bound, and gave a complete
description of the rotationally symmetric solutions.

There remains several further issues to be discussed.
Firstly, note that the identity (\ref{use}) and its gauged version
on HSS is essential for the existence of self-duality. In this respect,
it would be an intriguing problem to extend the above formalism 
to other non HSS coadjoint orbits, 
and also to higher non-compact group.
Quantization of the model is another problem to be addressed.
Secondly, it would be interesting to see whether
there exists  a  well-defined procedure in which
the non-relativistic NLSM of the generalized CS
Heisenberg ferromagnet system defined on the coadjoint orbits
\cite{oh2} could emerge as a non-relativistic limit
of the present relativistic NLSM. This will require in the course
a revelation of the connection between the symplectic
structure of HSS \cite{oh2} for the  non-relativistic NLSM 
and phase space structure of relativistic NLSM.

\vspace{0.3cm} 

I thank D. Chae, Y. Kim, K. Kimm, K. Lee, Q-H. Park,
 and C. Rim for useful discussions, Sung-Soo Kim for his 
invaluable help, and Prof. A. Strasburger for his hospitality
at XVI-th Workshop on Geometric Methods in Physics.
This work is supported in part 
by the Korea Science and Engineering Foundation 
through the project number
(95-0702-04-01-3), 
and  by the Ministry of Education through the
Research Institute for Basic Science  (BSRI/97-1419).

\newpage
\begin{figure}[bth]
\begin{center}
\epsfig{figure=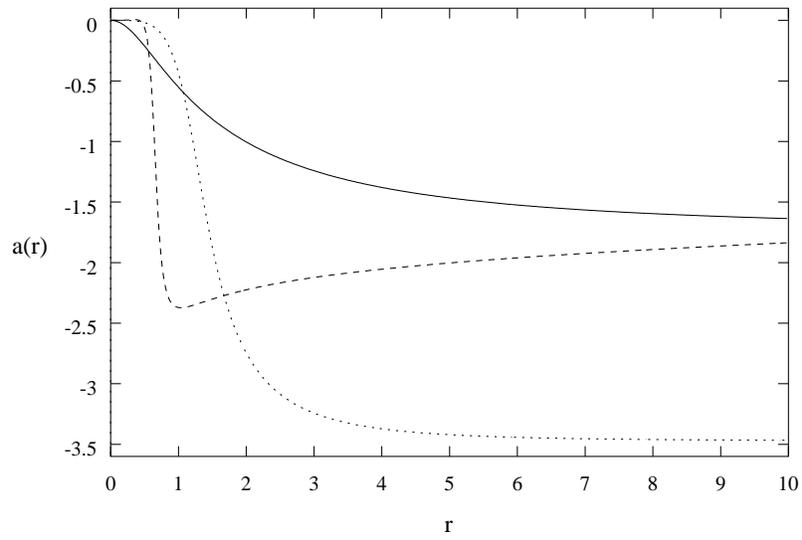,height=7cm}
\end{center}
\caption{$a(r)$ as a function of $r$ for $\kappa=1$ with
various $(n, V)$: $(0, 0)$ for solid line, $(-1, 0)$ for dotted line,
and $(-1, 1.5)$ for dashed line.}
\end{figure}
\newpage
\begin{figure}[tbh]
\begin{center}
\epsfig{figure=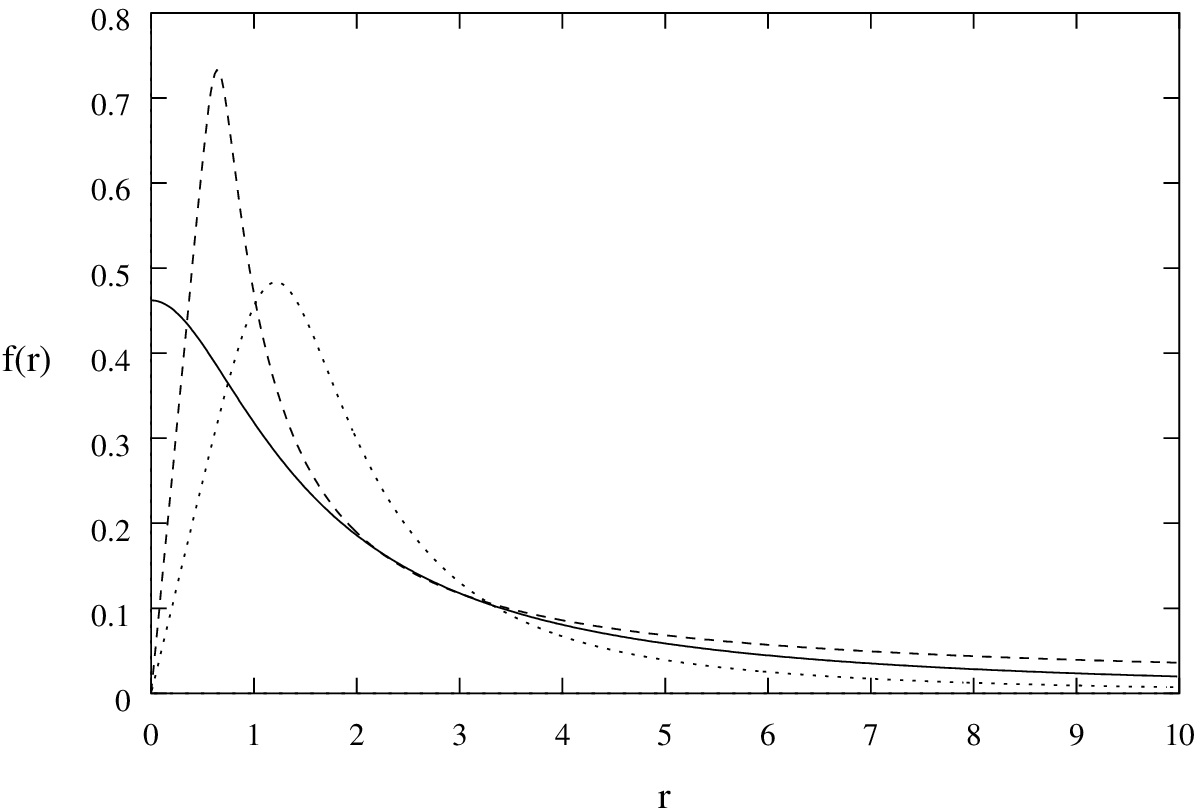,height=7cm}
\end{center}
\caption{$f(r)$ as a function of $r$ for $\kappa=1$ with
various $(n, V)$: $(0, 0)$ for solid line, $(-1, 0 )$ for
dotted line, and $(-1, 1.5)$ for dashed line.}
\end{figure}


\begin{thebibliography}{99}
\bibitem{note} We will only consider the simplest 
non-compact case $SU(1,1)$. Here, $\epsilon =-1$ is necessary
to render the energy to be non-negative 
in the representation in which the element $K$ is anti-Hermitian. 
Higher orbits of non-compact group needs more careful treatment.
\end{thebibliography}
\end{document}